\documentclass[iop]{emulateapj}

\pdfoutput=1 
\slugcomment{   }
\usepackage{amsmath} 
\usepackage{graphicx}
\usepackage{txfonts}
\usepackage{epsfig}
\usepackage{times}
\usepackage{natbib}
\bibpunct{(}{)}{,}{a}{}{}

\usepackage{tabularx}
\usepackage{amssymb}
\usepackage{balance}
\usepackage{natbib}
\bibpunct{(}{)}{;}{a}{}{,} 
\usepackage{color}
\usepackage{xspace}
\usepackage[normalem]{ulem} 
\usepackage[table]{xcolor} 
\usepackage{multirow}
\usepackage{rotating}
\usepackage{bbm}

\shorttitle{Activity and quiescence in Cl~J1449+0856 at $z=2$}
  
\shortauthors{Strazzullo et al.}

\begin{document}


\title{Deciphering the activity \emph{and} quiescence of high-redshift cluster environments:\\ ALMA observations \
of Cl~J1449+0856 at  \lowercase{$z=2$}}


\author{V. Strazzullo$^1$, R.~T. Coogan$^{2,3}$, E. Daddi$^3$,
  M.~T. Sargent$^2$, R. Gobat$^4$, F. Valentino$^5$, M. Bethermin$^6$,
  M. Pannella$^1$, M. Dickinson$^8$, A. Renzini$^9$,
  N. Arimoto$^{10}$, A. Cimatti$^{11}$, H. Dannerbauer$^{12,13}$,
  A. Finoguenov$^{14,15}$, D. Liu$^{16}$, M. Onodera$^{17,18}$}

\altaffiltext{1}{Faculty of Physics, Ludwig-Maximilians-Universit{\"a}t,
  Scheinerstr. 1, 81679 M{\"u}nchen, Germany -- vstrazz@usm.lmu.de}  
\altaffiltext{2}{Astronomy Centre, Department of Physics and Astronomy, University of Sussex, Brighton BN1 9QH, UK}
\altaffiltext{3}{CEA, IRFU, DAp, AIM, Universit\'e Paris-Saclay, Universit{\'e} Paris Diderot, Sorbonne Paris Cit\`e, CNRS, F-91191 Gif-sur-Yvette, France}
\altaffiltext{4}{Instituto de F{\'i}sica, Pontificia Universidad Cat{\'o}lica de Valpara{\'i}­so, Casilla 4059, Valpara{\'i}­so, Chile}
\altaffiltext{5}{Dark Cosmology Centre, Niels Bohr Institute, University of Copenhagen, Juliane Maries Vej 30, DK-2100 Copenhagen, Denmark}
\altaffiltext{6}{Aix Marseille Univ, CNRS, LAM, Laboratoire d'Astrophysique de Marseille, Marseille, France}
\altaffiltext{8}{National Optical Astronomy Observatory, 950 North Cherry A venue, Tucson, AZ  85719, USA}
\altaffiltext{9}{INAF-Osservatorio Astronomico di Padova, Vicolo dell'Osservatorio 5, I-35122, Padova, Italy}
\altaffiltext{10}{Astronomy Program, Department of Physics and Astronomy,
Seoul National University, 599 Gwanak-ro, Gwanak-gu, Seoul, 151-742, Korea}
\altaffiltext{11}{Dipartimento di Fisica e Astronomia, Universit{\'a} di Bologna, via Gobetti 93/2, I-40129, Bologna, Italy}
\altaffiltext{12}{Instituto de Astrofisica de Canarias (IAC), E-38205 La Laguna, Tenerife, Spain}
\altaffiltext{13}{Universidad de La Laguna, Dpto. Astrofisica, E-38206 La Laguna, Tenerife, Spain}
\altaffiltext{14}{Max-Planck-Institut f\"ur extraterrestrische Physik, Giessenbachstrasse 1, 85748 Garching, Germany}
\altaffiltext{15}{Department  of Physics, University of Helsinki, P.~O. Box 64, FI-00014,  Helsinki, Finland}
\altaffiltext{16}{Max Planck Institute for Astronomy, Konigstuhl 17, D-69117 Heidelberg, Germany}
\altaffiltext{17}{Subaru Telescope, National Astronomical Observatory of Japan, National Institutes of Natural Sciences (NINS), 650 North A'ohoku Place, Hilo, HI, 96720, USA}
\altaffiltext{18}{Department of Astronomical Science, SOKENDAI (The Graduate University for Advanced Studies), 650 North A'ohoku Place, Hilo, HI 96720, USA}

\begin{abstract}
We present {\it ALMA} observations of the 870$\mu$m continuum and
CO(4-3) line emission in the core of the galaxy cluster ClJ1449+0856
at $z=2$, a NIR-selected, X-ray detected system in the mass range of
typical progenitors of today's massive clusters. The 870$\mu$m map
reveals six F$_{870\mu m}>0.5$~mJy sources spread over an area of
$0.07$arcmin$^2$, giving an overdensity of a factor $\sim10$ (6) with
respect to blank field counts down to F$_{870\mu m}>1$~mJy
($>0.5$~mJy). On the other hand, deep CO(4-3) follow-up confirms
membership of three of these sources, but suggests that the remaining
three, including the brightest 870$\mu$m sources in the field
(F$_{870\mu m}\gtrsim2$~mJy), are likely interlopers.  The measurement
of 870$\mu$m continuum and CO(4-3) line fluxes at the positions of
previously-known cluster members provides a deep probe of dusty star
formation occurring in the core of this high-redshift structure,
adding up to a total SFR$\sim700\pm100$~M$_{\odot}$/yr and yielding an
integrated star formation rate density of
$\sim10^4$~M$_{\odot}$yr$^{-1}$Mpc$^{-3}$, five orders of magnitude
larger than in the field at the same epoch, due to the concentration
of star-forming galaxies in the small volume of the dense cluster
core.  The combination of these observations with previously available
{\it HST} imaging highlights the presence in this same volume of a
population of galaxies with already suppressed star formation. This
diverse composition of galaxy populations in ClJ1449+0856 is
especially highlighted at the very cluster center, where a complex
assembly of quiescent and star-forming sources is likely forming the
future Brightest Cluster Galaxy.
\end{abstract}



\section{Introduction} 
\label{sec:intro}
\setcounter{footnote}{0}

Thanks to a variety of efforts devoted to the search for distant
progenitors of galaxy clusters, pushing towards $z\sim2$ and beyond
\citep[e.g.][]{rosati1999,kurk2000,mullis2005,stanford2006,venemans2007,eisenhardt2008,wilson2009,andreon2009,papovich2010,santos2011,gobat2011,spitler2012,stanford2012,muzzin2013c,wylezalek2013,clements2014,bleem2015,strazzullo2015,casey2015,wang2016,flores-cacho2016,daddi2017,mantz2017,cai2017},
it has been possible in the last decade to significantly extend the
timeline and scope of galaxy evolution studies in the densest
high-redshift environments. This has eventually bridged the
historically divided cluster and proto-cluster regimes \citep[see
  recent review by][]{overzier2016}, at a cosmic time which is thought
- and indeed turns out - to be a transformational epoch for both
clusters and their galaxies. The synergy of observations at different
wavelengths, including optical/NIR to probe stellar populations and
galaxy structural properties, mid/far-IR to radio to probe star
formation rates, sub-mm for gas reservoirs, as well as X-ray, mid-IR
and radio for nuclear activity, has proved fundamental in exploring
the many facets of cluster galaxy populations, as discussed below.

From observations of massive cluster galaxies at lower redshifts
\citep[e.g.][]{andreon2006,depropris2007,lidman2008,mei2009,strazzullo2010b,mancone2010,wylezalek2014},
we expect that the epoch around $z\sim2$ corresponds to the transition
from a regime of widespread, high levels of star formation in dense
environments, to the quiescent regime characteristic of cluster cores
at $z\lesssim1$. Direct observations at high redshifts have in fact
detected increasing levels of star formation as well as nuclear and
merging activity in distant $z\gtrsim1.5$ groups and clusters
\citep[e.g.][and references
  therein]{hilton2010,hayashi2010,hayashi2011,zeimann2012,
  brodwin2013,lotz2013,
  dannerbauer2014,santos2015,popesso2015,ma2015,tran2015,wang2016,alberts2016,hung2016,krishnan2017,nantais2017}. At
the same time, passively evolving galaxies are often found to be
over-represented, to different degrees, in the densest regions of
these environments \citep[e.g.][]{steidel2005,kurk2009, papovich2010,
  gobat2013, strazzullo2013, tanaka2013,snyder2012,
  spitler2012,kubo2013,hatch2016}, and examples of (generally massive)
clusters with already very strongly suppressed star formation are also
found up to $z\sim2$
\citep[e.g.][]{strazzullo2010b,cooke2016,andreon2014,newman2014}.
Cluster selection, cluster-to-cluster variation, the intrinsically
transitional phase of galaxy populations at this time, and
observational difficulties, have all contributed to assemble a varied,
still unfinished picture that might at times still look controversial
in some aspects, and sometimes difficult to reconcile with theoretical
expectations \citep[e.g. ][]{granato2015}.

Thanks to expensive -- thus still limited to a relatively small number
of systems -- dedicated follow-up programs, a number of recent studies
have started investigating in more specific detail the properties of
both quiescent and star-forming galaxies in distant (proto-)cluster
environments. Such studies explored a variety of aspects including:
the environmental dependence of stellar ages and structure of passive
populations
\citep[e.g.][]{gobat2013,tanaka2013,strazzullo2013,newman2014,andreon2014,andreon2016,beifiori2017,prichard2017,lee-brown2017,chan2018};
the environmental dependence of the specific star formation rates,
metallicities, and dust attenuation properties of star-forming
galaxies
\citep[e.g.][]{tanaka2010,hatch2011,hayashi2011,hayashi2016,kulas2013,koyama2013,koyama2014,smail2014,cooke2014,tran2015,shimakawa2015,valentino2015,kacprzak2015,kewley2016,husband2016,shimakawa2018};
the environmental dependence of cold gas reservoirs fueling star
formation in dense environments
\citep[e.g.][]{aravena2012,emonts2013,casasola2013,ivison2013,tadaki2014,gullberg2016,casey2016,lee2017,noble2017,dannerbauer2017,rudnick2017,stach2017,hayashi2017,hayashi2018}. Results
from such investigations critically shape our understanding of galaxy
population properties - and of the processes affecting galaxy
evolution - in early dense environments, though the still very limited
cluster galaxy samples, small number of clusters probed, and selection
biases continue to preclude conclusive interpretations.

We present here new results from Atacama Large Millimetre Array ({\it
  ALMA}) observations of the galaxy cluster ClJ1449+0856
\citep[hereafter ClJ1449,][]{gobat2011} at $z=2$, complementing our
previous work on its galaxy populations
\citep{gobat2013,strazzullo2013,valentino2015,valentino2016,strazzullo2016}
with a critical independent vantage point. ClJ1449 was identified as
an overdensity of IRAC color selected galaxies, and spectroscopically
confirmed with now $\sim30$ spectroscopic members
\citep{gobat2011,gobat2013,valentino2015,coogan2018}. The estimated
halo mass based on its extended X-ray emission and stellar mass
content is $5-7 \times 10^{13}$~M$_{\odot}$ \citep[][and references
  therein]{valentino2016}, placing this structure in the mass range of
the average progenitors of today's typical massive clusters.  For what
directly concerns the results presented here, our previous work has
highlighted the mixed galaxy population in this cluster, consisting of
both quiescent and highly star-forming sources \citep[as well as
  AGNs,][]{gobat2013,strazzullo2013,valentino2015,valentino2016}. In
particular, in \citet[][hereafter S16]{strazzullo2016} we investigated
the nature of the massive red population characterising the cluster
core, in terms of dusty massive star-forming galaxies and sources with
already suppressed star formation, based on a purely photometric
analysis at optical/NIR wavelengths. On the other hand, in spite of
the statistical validity of this approach, the ultimate confirmation
of the nature of such sources, and an actual estimate of the
(obscured) star formation occurring in the cluster core, remains with
star formation rate indicators not biased by dust attenuation. ClJ1449
had been previously observed with {\it Spitzer}/MIPS (24$\mu$m) and
later also with {\it Herschel}/PACS and SPIRE, which indeed suggested
potentially high levels of star formation activity right in the
cluster core as already reported in the first study by
\citet{gobat2011}. However, the angular resolution and depth of these
data, and/or ambiguities with respect to contamination from nuclear
activity, hampered the effectiveness of these observations in
establishing a reliable picture of star formation and quenching in
this system.

In this work based on {\it ALMA} observations, we thus focus on three
main aspects: first, the quantification of star formation occurring in
the core of ClJ1449, for the first time using deep, high-resolution
star formation probes not biased by dust attenuation
(Sections~\ref{sec:ovd}, \ref{sec:dustySF}). Second, the constraints
set by these new observations on the first massive quiescent galaxies
that, even as early as $z\sim2$, are a significant feature of the core
of ClJ1449 (Section~\ref{sec:quiescent}). Finally, the direct
observation of the forming Brightest Cluster Galaxy (BCG,
Section~\ref{sec:bcg}), that especially thanks to the combination with
{\it Hubble Space Telescope} ({\it HST}) data produces a remarkably
detailed picture of a critical phase in the early evolution of
(proto-)BCGs. Besides the scope of this paper, the {\it ALMA}
observations discussed here enable the investigation of a
significantly wider range of questions related to the effect of the
environment on the properties and evolution of high-redshift cluster
galaxies: the companion paper by \citet[][hereafter C18]{coogan2018}
presents in particular the dust and gas properties of the {\it
  ALMA}-detected cluster members, and provides extensive descriptions
of all sub-mm and radio observations of ClJ1449.

We assume a $\Lambda$CDM cosmology with $\Omega_{M}$=0.3,
$\Omega_{\Lambda}$=0.7, H$_{0}$=70~km~s$^{-1}$~Mpc$^{-1}$.  Stellar
masses and star formation rates are quoted for a \citet{salpeter1955}
IMF. Magnitudes and colors are quoted in the AB system.

\begin{figure*}[ht!]
\begin{center}
 \includegraphics[width=0.99\textwidth,bb= 113 248 845 609, clip]{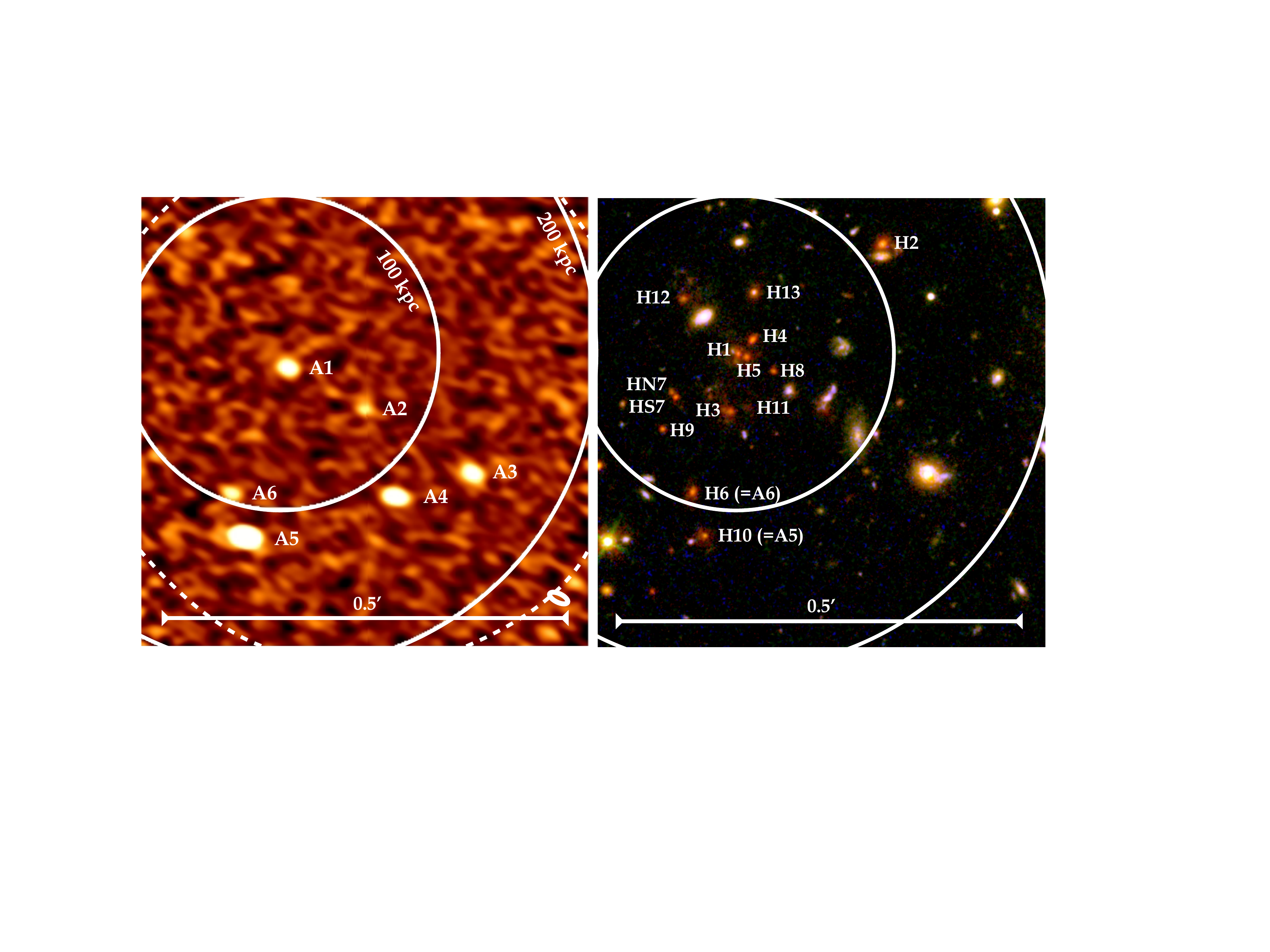}
\end{center}
\caption{The {\it ALMA} 870$\mu$m map (left) and {\it HST}
  F606W-F105W-F140W pseudo-color image (right) of the same region in
  the core of ClJ1449. Sources discussed in this work are
  labeled. Note the different identifications for the {\it
    ALMA}/870$\mu$m sources (IDs~A1 to A6) and {\it HST}/F140W sources
  from S16 (IDs~H1 to H13, corresponding to IDs~1 to 13 in S16). The
  field shown is $\sim$0.56' on a side, $\sim$290~kpc (proper) at the
  cluster redshift. White solid circles show clustercentric distances
  of 100 and 200~kpc as indicated (the estimated cluster virial radius
  is $\sim$400~kpc). The white dashed circle shows the primary beam
  FWHM of the {\it ALMA} band 4 observations.
\label{fig:870map} }
\end{figure*}

\section{Observations}
\label{sec:obs}

The central area of ClJ1449 was observed in {\it ALMA} Cycle~1 and 3
programs 2012.1.00885.S and 2015.1.01355.S.  The Cycle~1 program
obtained a band 7 mosaic probing 870$\mu$m continuum over a $\sim
0.3$arcmin$^2$ region in the cluster central area. Observations were
completed in December 2014 for a total on-source time of $\sim2.3$h.
The probed field, offset by 8'' from the cluster center, reaches out
to clustercentric distances $\sim$100-200~kpc depending on the
direction (see Figure~\ref{fig:870map}). For comparison, the estimated
cluster virial radius is $r_{200}\sim0.4\pm0.1$~Mpc
\citep{gobat2013,valentino2016}.

The synthesised beam of
FWHM$\sim 1.4$''$\times 0.6$'' is well suited for the size of $z\sim2$
galaxies and to avoid confusion in the crowded cluster-core
environment. The map has an rms sensitivity of $\sim$70~$\mu$Jy/beam
and thus, as a reference, reaches down to a star formation rate (SFR)
of $\sim$40~M$_{\odot}$/yr \citep[3$\sigma$,][see
  Sec.~\ref{sec:dustySF}]{bethermin2012} for ``main sequence''
\citep[MS,][]{elbaz2011} star-forming galaxies at $z\sim2$,
corresponding to stellar masses $\sim2\times10^{10}$~M$_{\odot}$
\citep{sargent2014}. The same program also obtained a single
overlapping pointing in band 3 to probe CO(3-2) emission from cluster
galaxies at matched depth with the 870$\mu$m continuum, with the main
goals of confirming cluster membership of sources detected in the
870$\mu$m map, and estimating gas reservoirs fueling their star
formation. However, as the later aquired CO(4-3) observations
discussed below are deeper (see C18), and as higher-order CO
transitions are better SFR (rather than total molecular gas mass)
tracers \citep[e.g.]{daddi2015,liu2015}, these CO(3-2) data are not
used for the purposes of this work. 

The Cycle~3 follow-up observed again this region with a band 4
pointing to probe CO(4-3) emission from cluster members. Observations
were completed in May 2016, for a total on-source time of $\sim2$h.
The FWHM of the primary beam is $\sim$41'', well matched to the
870$\mu$m mosaic (Fig.~\ref{fig:870map}). The synthesised beam FWHM is
$\sim 1.2$''$\times 1$''. The rms is 10~mJy~km/s over 100~km/s,
corresponding to a $3\sigma$ detection limit of $\sim$35~M$_{\odot}$/yr
for a MS galaxy at $z=2$ with a CO(4-3) line FWHM of $\sim$400~km/s.

We focus here on the constraints on star formation activity and its
suppression in the core region of ClJ1449 derived from the
observations of 870$\mu$m continuum and CO(4-3) line emission. We
refer the reader to C18 for a full, extensive summary of the {\it
  ALMA} observations, and for a detailed description of the
measurements of the 870$\mu$m continuum and CO(4-3) line fluxes that
we use in this work.  Summarising those aspects most important to the
analysis presented here, we note that C18 detected continuum sources
and independently searched for spectral lines in the band 7 and band 4
observations. Then, both 870$\mu$m continuum and CO(4-3) line fluxes
were measured at the positions of all these mm-detected sources, as
well as at the positions of all known cluster members from our
previous optical/NIR studies
\citep{gobat2011,gobat2013,strazzullo2013,valentino2015,strazzullo2016}. All
the analysis and results below are based on these measurements as
described in C18.

\begin{figure*}[t]
\begin{center}
 \includegraphics[width=1.01\textwidth,bb=2 610 708 767,clip]{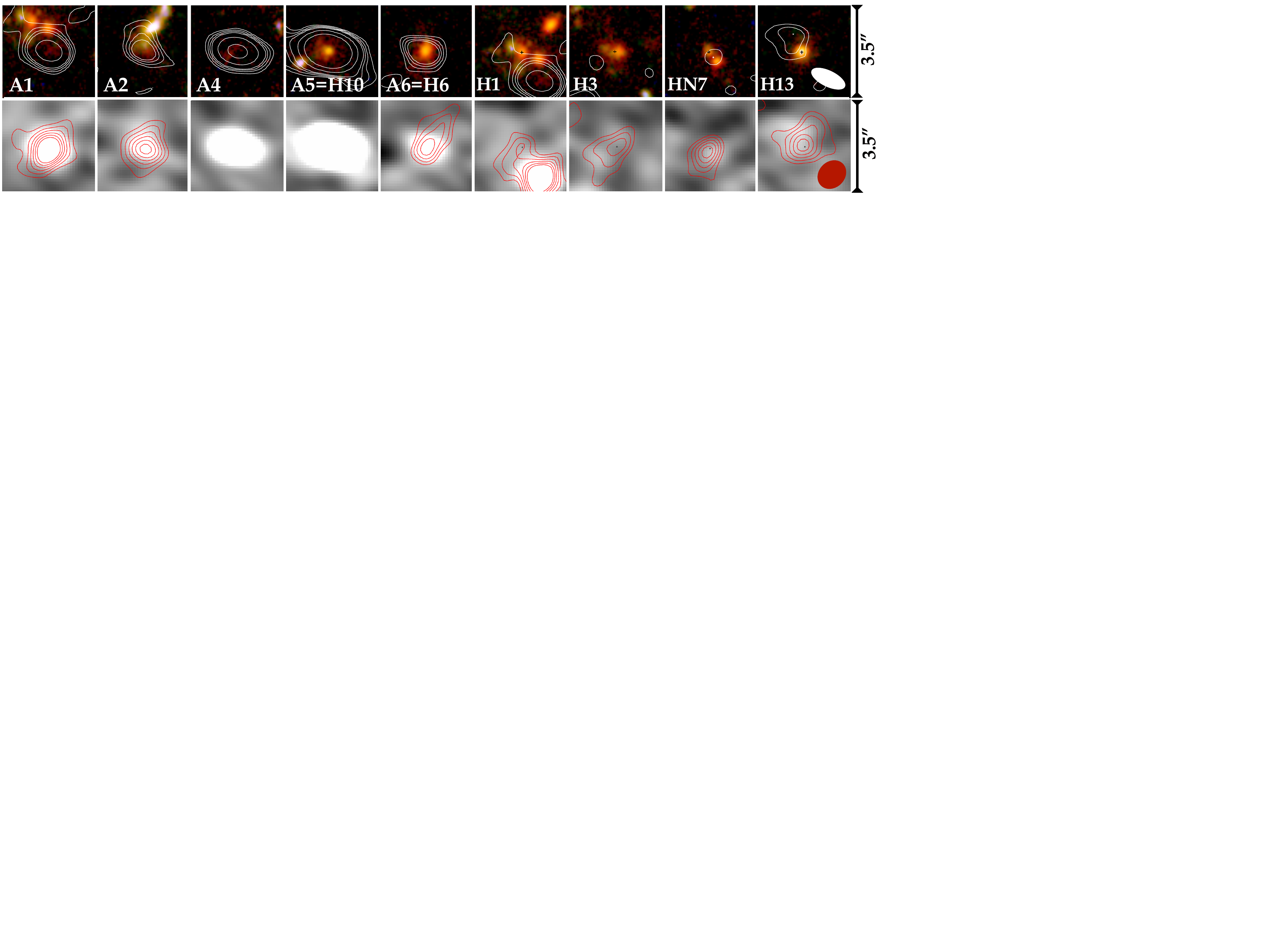}
\end{center}
\caption{{\it Top:} {\it HST} pseudo-color image cutouts (F606W-F105W-F140W)
  of sources discussed in this work (see labeling in
  Fig.~\ref{fig:870map}), with overlaid (white) 870$\mu$m contours
  ($2,3,4,5,10, 20\sigma$). {\it Bottom:} 870$\mu$m cutouts with
  overlaid (red) CO(4-3) contours ($2,3,4,5,6,7\sigma$). For
  IDs~$H1,H3,HN7,H13$ black crosses mark the {\it HST}/F140W galaxy
  position. White and red ellipses in rightmost panels show the
  symthesised beam of the 870$\mu$m and band~4 observations, respectively.
\label{fig:cutouts}}
\end{figure*}

\section{Overdensity  of \lowercase{870$\mu$m} sources}
\label{sec:ovd}

The $870\mu$m map of the core of ClJ1449 reveals six continuum sources
($S/N>4$) with F$_{870\mu m}>0.5$~mJy (Figure~\ref{fig:870map}),
giving a projected source density in the $\sim0.3$arcmin$^2$ survey
field that is a factor $\sim6$ higher than expected from blank-field
counts \citep[e.g.~$\sim1.0\pm0.8$ sources would be expected
  from][]{oteo2016}. In fact, all six sources are concentrated within
a circle of $r\sim0.15'$.
Four of the six 870$\mu$m sources are brighter than 1~mJy, resulting
in a projected overdensity of $\sim10$ \citep[$\sim0.3\pm0.3$ sources
  expected to this flux limit from][]{oteo2016}.
We note that, as discussed below, the two brightest $870\mu$m sources
are likely to be background interlopers. We estimate their flux
magnification due to the cluster potential, assuming a spherical NFW
halo of mass $6\times10^{13}$~M$_{\odot}$ and a concentration in the
range $c=1-5$. For the brighest, $\sim$6~mJy source (labeled A5
below), the estimated photometric redshift $z\sim2.8$ yields a
magnification factor of $\sim12\%$ (5-18\% within the $3\sigma$ range
of the photometric redshift). For the other source (A4,
$\sim$1.9~mJy), we estimate a magnification of about 9\% to 20\%
(30\%, 40\%) for a source redshift from 2.5 to 3.5 (5, 7). Even for a
source redshift $z\sim7$, the flux would still be brighter than
1~mJy. We thus conclude that the overdensity of mm-bright sources is
not likely significantly affected by lensing magnification by the
cluster potential.

Figure ~\ref{fig:870map} (left) identifies all {\it ALMA}-detected
sources discussed in this work (labeled A1 to A6), while the
right-hand panel identifies the {\it HST}-detected (F140W) sources
(labeled H1 to H13, note that H6 and A6 - and H10 and A5 - correspond
to the same galaxy, as indicated).  Of the six 870$\mu$m sources, two
have spectroscopic redshifts measured in the \citet{gobat2013} {\it
  HST} grism follow-up of ClJ1449: a foreground galaxy at $z\sim1.3$
(A3, thus ignored henceforth), and the cluster member A6=H6. None of
the other 870$\mu$m sources has an optical/NIR spectroscopic redshift
determination. The {\it HST} counterparts to A1, A2 and A4 are very
faint\footnote{A1 and A2 are also close to a bright neighbour and were
  not extracted as individual sources in our F140W-based catalogs
  \citep[e.g.,][]{strazzullo2013} until {\it HST}/F105W-band imaging
  (S16) and {\it ALMA} observations pointed to these sources being
  separate components.}, and were thus not included in our previous
studies of galaxy populations in ClJ1449. A5 was included in our
previous work and was deemed to be likely an interloper at
z$_{phot}\sim2.8$ \citep{strazzullo2013}.

On the other hand, A1 and A2 both show a highly significant detection
of CO(4-3) line emission (Fig.~\ref{fig:cutouts} and C18), securely
confirming their cluster membership. However, no lines are detected
for the two brightest 870$\mu$m sources, A4 and A5.  In fact, as
discussed at length in C18, in spite of their high 870$\mu$m fluxes no
lines are detected for these sources in {\it any} of our data sets
probing CO(4-3), CO(3-2) and CO(1-0) at the cluster redshift, as well
as bright mm lines ([CI](2-1), CII, CO transitions up to CO(7-6)) over
a significant fraction of the $1<z<9$ range. Nonetheless, several
redshifts remain unprobed, notably including the range around the
photometric redshift of A5 (C18). Therefore we do not presently have
confirmation of the redshift of A4 and A5. We note that the likelihood
of observing two such bright sources unrelated with ClJ1449 in the
small field probed is extremely low: as discussed in more detail in
C18, these sources might in principle still be cluster members with
very recently and rapidly suppressed star formation, with the lack of
CO(4-3) emission being potentially reconciled with their bright
870$\mu$m continuum by the gas and dust tracing star formation on
different timescales. Nonetheless, given their large 870$\mu$m fluxes
and thus expected very bright CO line emission compared to the depth
of our observations (Fig.~\ref{fig:870vsco}), we currently conclude
that, at face value, the most likely explanation is that A4 and A5 are
interlopers. Among the six bright 870$\mu$m sources, only A1, A2 and
A6 are thus confirmed to belong to the cluster.

\section{Activity and quiescence in ClJ1449}

\subsection{Dusty star formation in the cluster core}
\label{sec:dustySF}

\begin{figure}[ht!]
\begin{center}
 \includegraphics[width=0.49\textwidth,bb=68 368 476 707,clip]{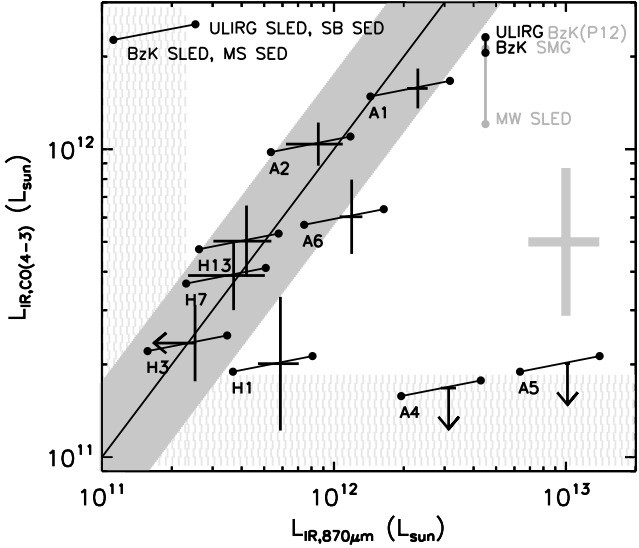}
\caption{The IR luminosities of mm-detected sources as estimated from
  the 870$\mu$m continuum fluxes (L$_{IR,870\mu m}$) and from the
  CO(4-3) line emission (L$_{IR,CO43}$). The double symbol shows
  L$_{IR,CO43}$ as estimated assuming a BzK or ULIRG CO SLED, and
  L$_{IR,870\mu m}$ as estimated with a main sequence (MS) or
  starburst (SB) SED, as indicated. For further comparison, the
  systematics on L$_{IR,CO43}$ corresponding to a range of different
  assumptions on the CO SLED are shown in the top right corner (see
  text).  Black error bars plotted in the middle of the two estimates
  show the uncertainties due to the 870$\mu$m and CO(4-3) flux
  measurements. The black line marks the bisector (L$_{IR,870\mu m}$ =
  L$_{IR,CO43}$).  The thick gray band along the bisector and the gray
  cross both show the uncertainty deriving from the ($1\sigma$)
  scatter in the adopted scaling relations (see text). Arrows show
  $2\sigma$ upper limits (see text). Hatched areas are below the
  $3\sigma$ limits for MS/BzK sources (see text). Note the clear
  outliers A4 and A5, which we therefore conclude are likely
  interlopers.
\label{fig:870vsco}}
\end{center}
\end{figure}

CO(4-3) emission is detected at $>3\sigma$ for a total of seven
cluster galaxies, including previously-known members (H1, H3, H6=A6,
HN7, H13) and those newly confirmed by the detection of the CO line
itself (A1, A2, Sec.~\ref{sec:ovd}). All but one of these (H3) also
have a $>2.5\sigma$ 870$\mu$m detection (Figures~\ref{fig:cutouts},
\ref{fig:870vsco}). Figure \ref{fig:870vsco} shows the infrared (IR)
luminosities L$_{IR}$ of these sources as estimated from the 870$\mu$m
continuum flux (L$_{IR,870\mu m}$) or the CO(4-3) line
flux\footnote{We use CO(4-3) line fluxes corrected for flux boosting
  as detailed in C18.}  (L$_{IR,CO43}$).

The IR luminosity estimate L$_{IR,870\mu m}$ was derived from the
measured 870$\mu$m flux using the average MS and starburst (SB) SEDs
from \citet{bethermin2012} at $z=2$.  The L$_{IR,CO43}$ estimate was
derived from the measured CO(4-3) line flux by assuming the CO SLEDs
(in particular, the CO(5-4)/CO(4-3) line ratio) of
ULIRGs\footnote{Given the significant uncertainties on the average
  CO(5-4) and CO(4-3) line fluxes adopted for ULIRGs \citep{daddi2015}, the  CO(5-4)/CO(4-3) line
  ratio used here is based on a weighted fit of the ULIRGs SLED from CO(3-2) to CO(5-4)
  rather than on the actual ratio of CO(5-4) and CO(4-3) line
  fluxes. As discussed, this ULIRGs line ratio is shown - for comparison with the BzK line ratio -  as an indication of the impact
  of the adopted CO SLED, and the exact value adopted does not have
  significant effects on the results of this work.}  (intended to
represent starbursts), and star-forming BzK galaxies \citep[intended
  to represent MS sources,
  from][respectively]{papadopoulos2012,daddi2015}, and the CO(5-4)
vs.\ L$_{IR}$ relation from \citet{daddi2015}.  We consider the
adopted assumptions on the CO(5-4)/CO(4-3) line ratio as the most
appropriate for galaxies in this sample, but we also show for
comparison in Fig.~\ref{fig:870vsco} the effect on the L$_{IR,CO43}$
estimate of a range of differents assumptions on the CO SLED,
including those measured for the Milky Way \citep[inner
  region,][]{fixsen1999}, SMGs \citep{bothwell2013}, and the
\citet{papadopoulos2012} model \citep[see discussion in][and
  references therein]{daddi2015}. The CO(5-4)/CO(4-3) line ratios from
these different determinations are affected to different degrees by
measurement uncertainties (see the original papers for details), but
for what concerns this work we note that the impact of the different
plausible line ratios (including their uncertainties) on our
L$_{IR,CO43}$ estimate is clearly sub-dominant with respect to the
scatter in the CO(5-4) vs.\ L$_{IR}$ relation, as well as to the
typical measurement errors of CO(4-3) line fluxes in this work.

Figure~\ref{fig:870vsco} shows these L$_{IR,870\mu m}$ and
L$_{IR,CO43}$ estimates and the related uncertainties as follows.
Black error bars show statistical uncertainties from flux measurement
errors. The double symbol adopted for all sources highlights the
systematic uncertainties in estimating IR luminosities from the
870$\mu$m continuum assuming a MS or SB SED, or from CO(4-3) line
fluxes assuming the BzK or ULIRG SLED, as indicated. The thick gray
error bar and shaded area along the bisector show the estimated
intrinsic scatter of the adopted scaling relations, that is the
scatter in SED shape (dust temperature) and in the CO(5-4)
vs.\ L$_{IR}$ relation (concerning L$_{IR,870\mu m}$ and
L$_{IR,CO43}$, respectively). Hatched regions in the figure mark IR
luminosities below a corresponding reference $3\sigma$ limit estimated
by assuming: for L$_{IR,870\mu m}$, the $3\sigma$ limit of the
870$\mu$m map (Sec.~\ref{sec:obs}) and the MS SED; for L$_{IR,CO43}$,
the $3\sigma$ limit of the band~4 observations at field center
(Sec.~\ref{sec:obs}), a line width of 400 km/s, the BzK
CO(5-4)/CO(4-3) line ratio, and the CO(5-4) vs.\ L$_{IR}$ relation.
While these are shown as an indication, the measurements, errors and
upper limits shown for the individual sources account for their actual
position within the band~4 primary beam FWHM, and highlight the
systematics due to the SED or CO SLED choice as discussed, though
upper limits for L$_{IR,CO43}$ still assume a line width of 400~km/s.

As Figure~\ref{fig:870vsco} shows, the L$_{IR,CO43}$ and L$_{IR,
  870\mu m}$ IR luminosity estimates are typically consistent within
the estimated uncertainties. The two obvious exceptions are A4 and A5,
which both have high L$_{IR, 870\mu m}$ from the bright 870$\mu$m flux
but no CO emission, leading to an inconsistent upper limit on
L$_{IR,CO43}$ even when accounting for the estimated uncertainties. As
discussed above, we therefore conclude that these sources are in fact
interlopers.  For all other sources, the consistency of L$_{IR,870\mu
  m}$ and L$_{IR,CO43}$, besides ensuring cluster membership of the
870$\mu$m detections, also confirms the reliability of the SFR
estimates.

Summing the derived infrared luminosities of cluster members within
the probed $\sim0.08$~Mpc$^2$ (proper, at z=2) region yields a total
L$_{IR}\sim4.3\pm0.5\times10^{12}$L$_{\odot}$ (the error corresponding
to the range obtained from L$_{IR,CO43}$ with both ULIRG and BzK
SLEDs, and L$_{IR,870\mu m}$ with a MS SED\footnote{Up to
  $\sim8\times10^{12}$L$_{\odot}$ from L$_{IR,870\mu m}$ if assuming a
  SB SED for all sources, corresponding to a total SFR of
  $\sim1400$~M$_{\odot}$/yr.}), corresponding to a total SFR of
$\sim700\pm100$~M$_{\odot}$/yr (adopting the \citet{kennicutt1998}
calibration).  This yields an overall projected SFR density of
$\sim0.9\pm0.1\times10^{4}$~M$_{\odot}$~yr$^{-1}$Mpc$^{-2}$, and a SFR
volume density
$\sim1.0\pm0.1\times10^{4}$~M$_{\odot}$~yr$^{-1}$Mpc$^{-3}$ within the
probed region\footnote{Up to
  $\sim1.7\times10^{4}$~M$_{\odot}$~yr$^{-1}$Mpc$^{-2}$ and
  $\sim2\times10^{4}$~M$_{\odot}$~yr$^{-1}$Mpc$^{-3}$ from
  L$_{IR,870\mu m}$ if assuming a SB SED for all sources.} (over the
probed fraction of the virial volume, given the estimated cluster
virial radius and assuming that the cluster is spherical). Again,
these estimates assume that the two brightest 870$\mu$m sources A4 and
A5 are interlopers: A5 for itself would otherwise contribute a
SFR$\sim$1000~M$_{\odot}$/yr.

The total unobscured SFR of the {\it ALMA}--detected cluster members
as estimated from the rest-frame UV luminosity L$_{UV}$ is
$<20$~M$_{\odot}$/yr. Given the SFR threshold reached by these
observations, the high L$_{IR}$/L$_{UV}$ of the resulting {\it
  ALMA}--detected sample further highlights how galaxy populations in
this cluster core are unusually skewed towards very reddened sources
(see also Fig.~\ref{fig:cutouts}, further discussion in Sec.~\ref{sec:colordist} and in C18).  For
comparison, the total unobscured SFR of all cluster galaxies within
the same region is estimated to be in the range
$100\pm20$~M$_{\odot}$/yr, after correcting for incompleteness using
the field UV luminosity functions from \citet{parsa2016,alavi2016}.

The measured SFR density is obviously orders of magnitude higher than
the field average at the same redshift
\citep[e.g.][]{madaudickinson2014}, as observed in various kinds of
other high-redshift structures
\citep[e.g.][]{clements2014,dannerbauer2014,tran2015,santos2015,wang2016}. This
is in fact largely due to the concentration of (star-forming) galaxies
within the small volume of the dense cluster core, rather than to
individual galaxies having particularly high SFRs. In fact, the
overall specific SFR in the {\it ALMA}-probed field is
sSFR$\sim1.1\pm0.6$~Gyr$^{-1}$, compared to a field average at $z=2$
of $\sim1.9^{+1.9}_{-0.9}$~Gyr$^{-1}$ \citep{madaudickinson2014}.  As
shown below (Fig.\ref{fig:almaRS}) and also discussed in C18, the SFRs
of individual sources are generally consistent with MS levels, with
the possible exception of the two brightest sources A1 and A2 having
higher SFRs.

By comparison with the cluster mass
M$_{halo}\sim5-7\times10^{13}$~M$_{\odot}$, the estimated
L$_{IR}$-derived SFR within the probed volume gives a lower limit
(that is, not correcting for the part of the virial volume left
unprobed by our {\it ALMA} observations) to the total SFR density
SFR/M$_{halo}$ of $\sim$1300$\pm$400~M$_{\odot}$~yr$^{-1}$ /
10$^{14}$~M$_{\odot}$, after a small correction of the total IR
luminosity for the $>3\sigma$ L$_{IR,870\mu m}$ sample down to
$L_{IR}=10^7$L$_{\odot}$ assuming the \citet{popesso2015a} group LF at
$z\sim1.6$, which should be the most appropriate for this system
\citep[see discussion in][]{popesso2015a}. At face value, this is in
line with the \citet{popesso2015} prediction at $z=2$ for massive
groups, although we remind that this is a lower limit and it is
currently not possible to reliably estimate the overall contribution
of the cluster outskirts (we note though that, in the available
observations, essentially all the measured IR luminosity is
contributed by the very central cluster region, $r\lesssim100$~kpc,
see Fig.~\ref{fig:870map}). The derived lower limit to SFR/M$_{halo}$
lies at the upper edge of the \citet{alberts2016} measurements at
$z\sim1.4$ (accounting for the different IMF and marginal correction
to the same $L_{IR}$ limit), in agreement with the expected further
increase out to $z=2$, though we also note that the
\citet{alberts2016} clusters have larger estimated halo masses in the
range $2-5\times10^{14}$~M$_{\odot}$, thus are expected to have lower
SFR/M$_{halo}$ \citep[e.g.][]{webb2013,popesso2015}. Indeed, results
from the lower halo mass sample of \citet{alberts2014} would give
significantly higher SFR/M$_{halo}$ in the same redshift range
\citep[see discussion in][]{alberts2016}, also higher than our lower
limit measured here. Similarly, although our lower limit tends to be
higher than measurements by e.g., \citet{smail2014,ma2015,santos2015}
on clusters of very different masses ($8\times10^{13}$~M$_{\odot}$ to
$5\times10^{14}$M$_\odot$) at $z\sim1.5-1.6$ (SFR/M$_{halo}$ overall
in the range $\sim500-1000$~M$_{\odot}$~yr$^{-1}$ /
10$^{14}$~M$_{\odot}$), it would be fully in line with these
measurements for a SFR density evolution similar to what predicted by
e.g. \citet{geach2006}. For comparison, the
$\sim3400$~M$_{\odot}$~yr$^{-1}$ observed within the 80~kpc core of the
similarly massive (M$_{halo}\sim8\times10^{13}$~M$_{\odot}$)
\citet{wang2016} cluster at $z=2.5$, result in a lower limit
SFR/M$_{halo}>4000$~M$_{\odot}$~yr$^{-1}$ / 10$^{14}$~M$_{\odot}$.

\subsubsection{Color distribution of {\it ALMA}-detected cluster galaxies}
\label{sec:colordist}

\begin{figure}
\begin{center}
  \includegraphics[width=0.45\textwidth,bb=82 405 539 700 ,clip  ]{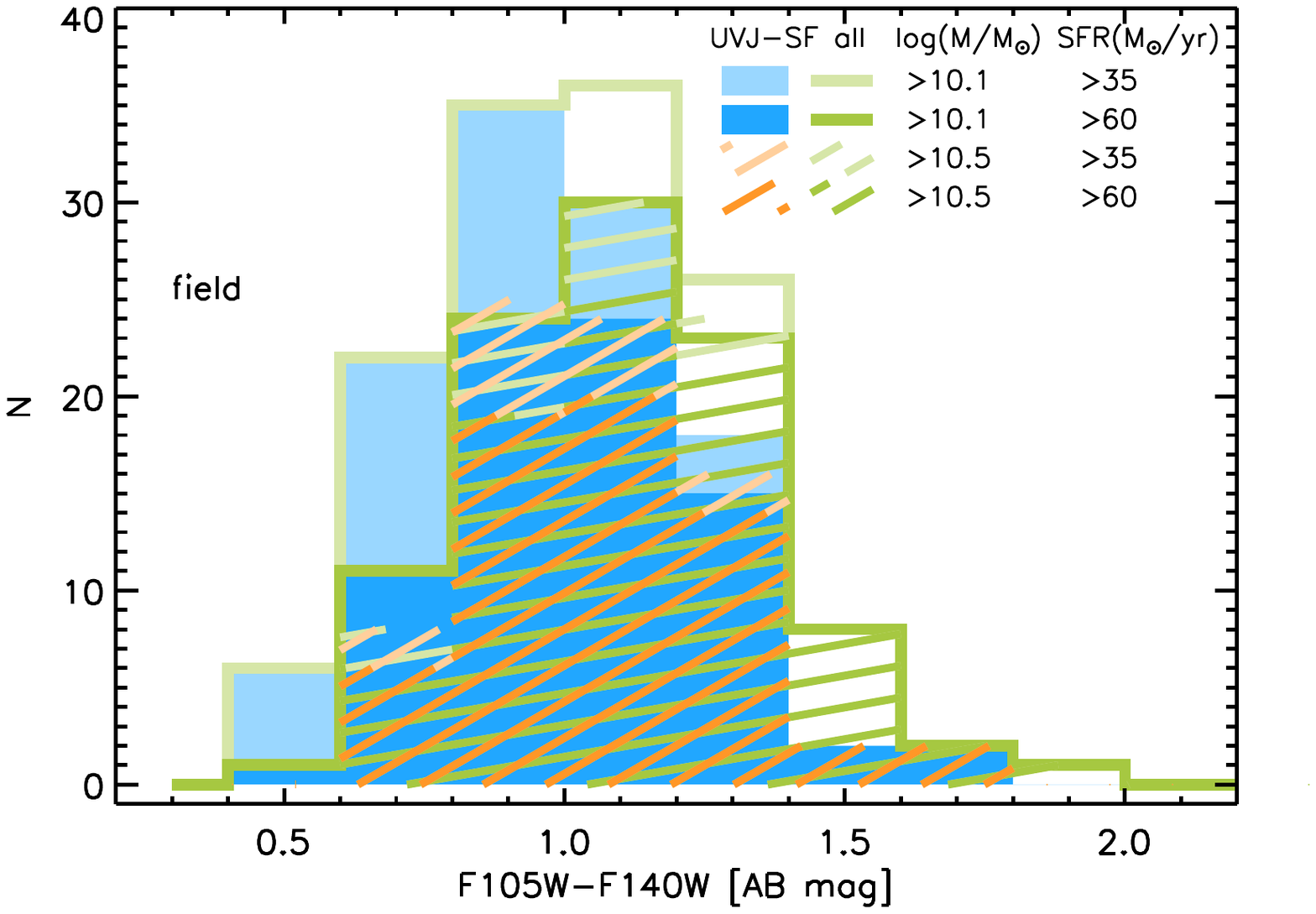}
  \includegraphics[width=0.45\textwidth,bb=82 370 539 700 ,clip  ]{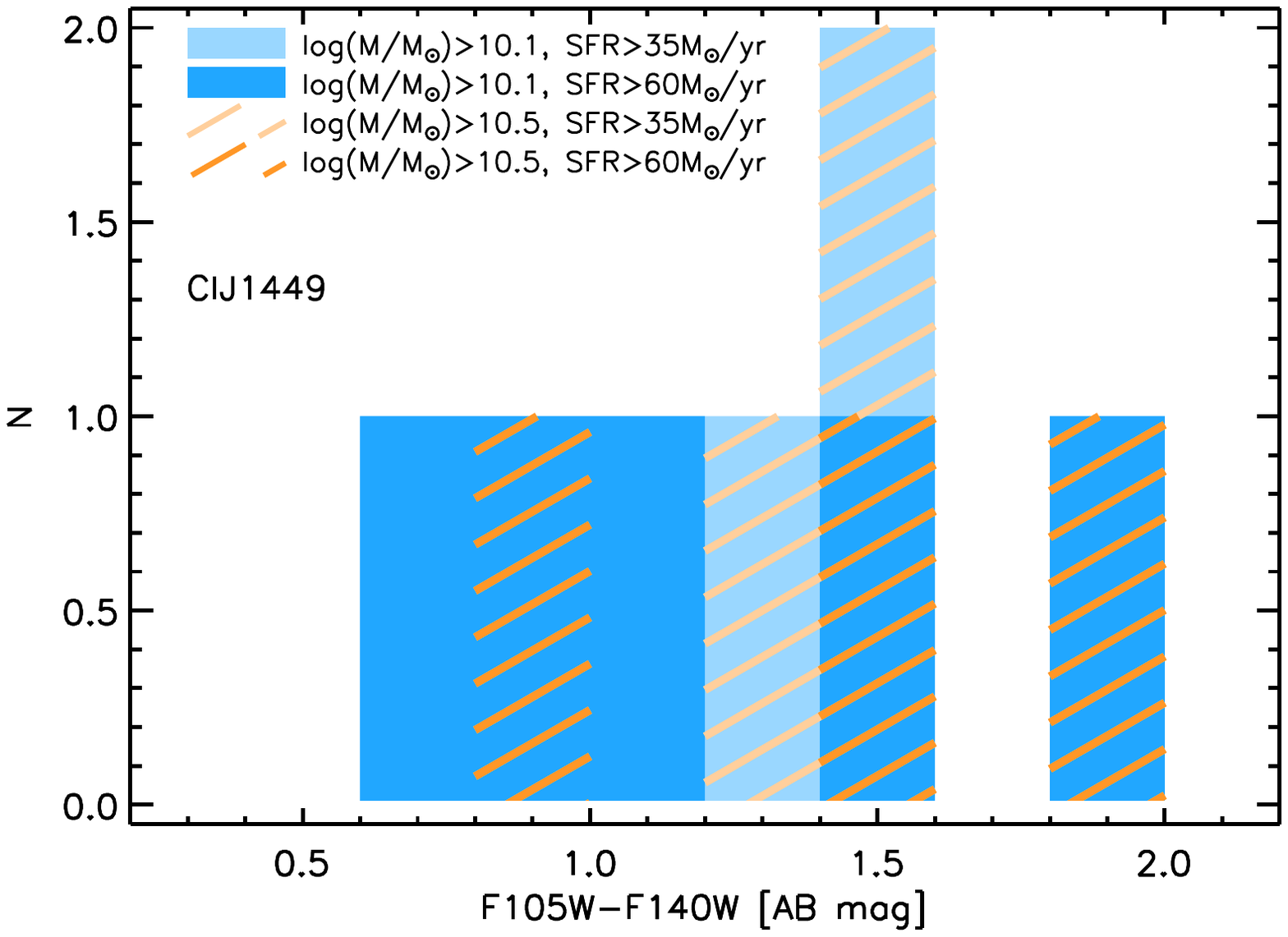}
\end{center}
\caption{The color distribution of stellar mass and SFR-selected
  samples (with different thresholds as indicated) in a control field
  (GOODS-S, top panel) and in the cluster field probed by
  {\it ALMA} observations (bottom panel). Note that SFRs for cluster members
  are based on CO(4-3) line emission, while the SFRs for the field
  comparison samples are derived from SED fitting.  The green and blue/orange
  histograms in the top panel refer to field samples including all
  galaxies and only UVJ-star-forming sources, respectively (see text).
  \label{fig:sfrvscolors}}
\end{figure}

\begin{figure*}[]
\begin{center}
\includegraphics[height=0.3\textheight,bb=67 356 462 719,clip]{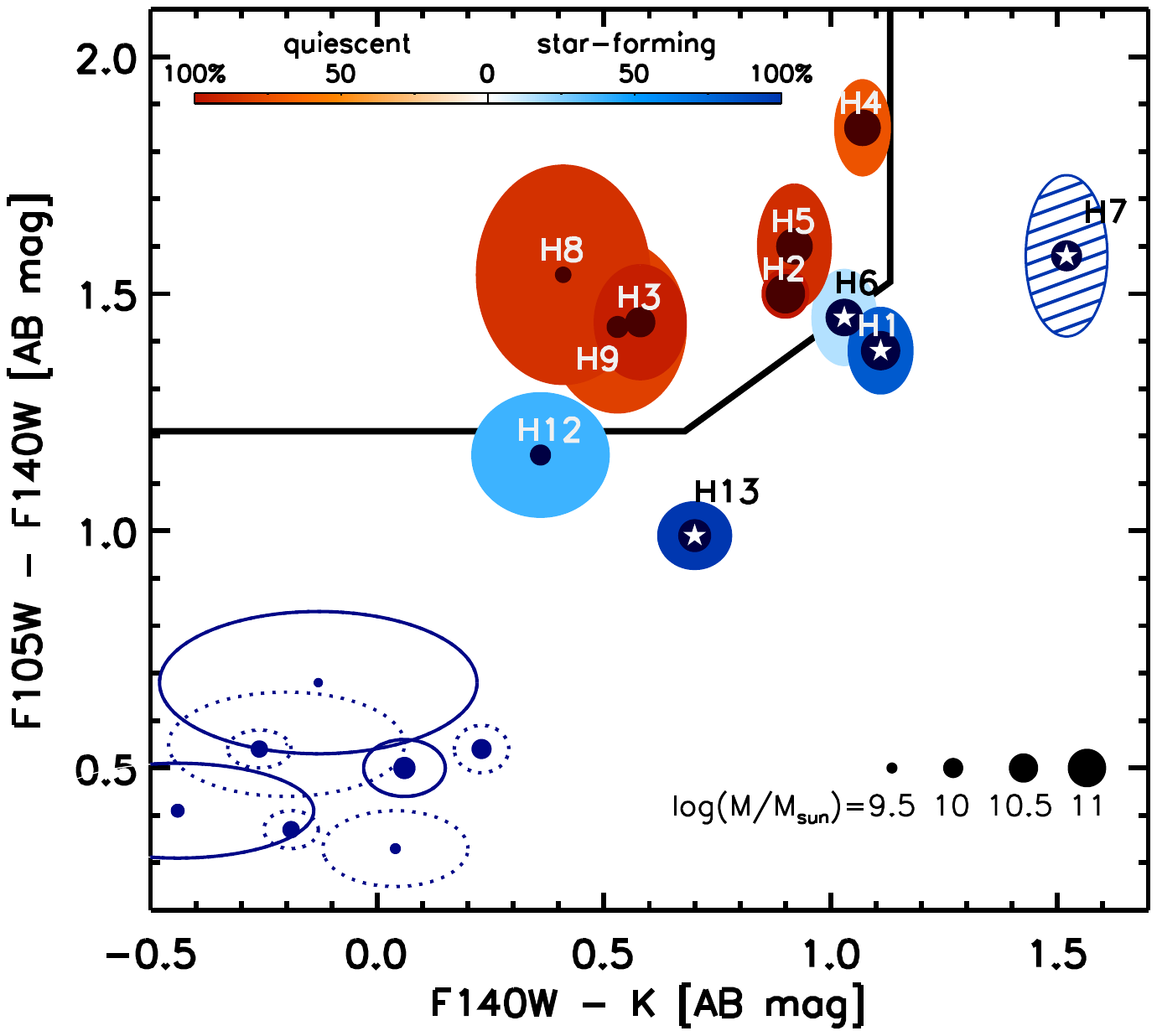}%
\hspace{.5cm}
\includegraphics[height=0.3\textheight,bb=56 360 458 690,clip]{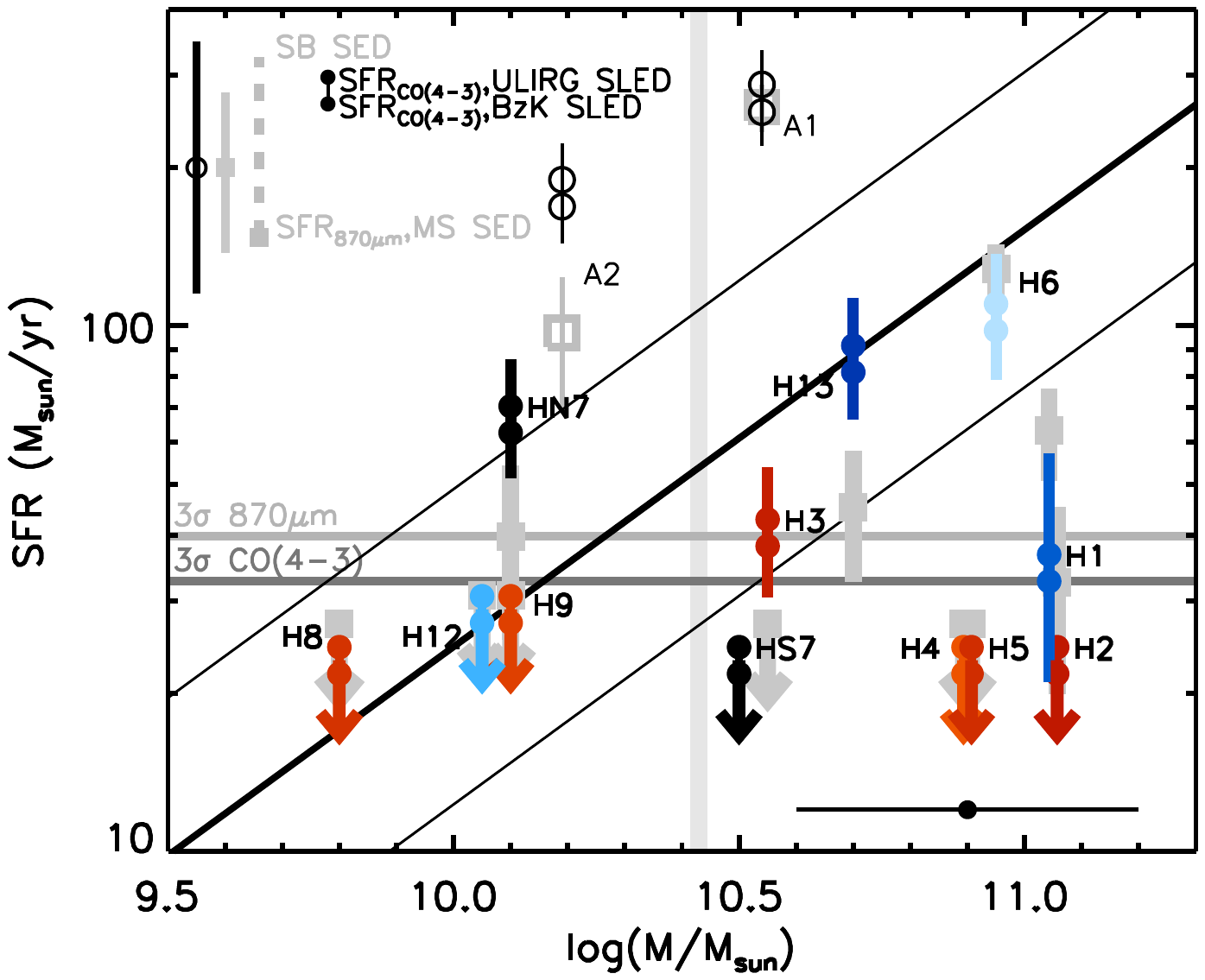}
\end{center}
\caption{The {\it ALMA} view of red cluster galaxies. {\it Left:} the
  color classification (F140W-K vs./ F105W-F140W) of the red galaxy
  sample as defined in S16. Galaxies are shown with blue or red
  circles (circle size scales with stellar mass as indicated)
  according to the classification as star-forming or quiescent,
  respectively. The (Gaussian) probability of the given classification
  for each source (see color bar) is defined from photometric
  uncertainties (shown as ellipses), and does not include
  uncertainties on the definition of the color selection. Filled
  labeled ellipses show galaxies in S16's red sample. H7 shows the
  combined colors of HN7 and HS7 (see text). White stars mark
  870$\mu$m-detected sources.  For reference, blue spectroscopic
  cluster members are also shown as empty ellipses (solid or dotted
  for sources in or outside the {\it ALMA}~870$\mu$m field). {\it
    Right:} The stellar mass vs.\ SFR as determined from L$_{IR,CO43}$
  (circles). Color-coding in shades of red (blue) reflects the
  reliability of the classification as quiescent (star-forming)
  according to the color bar in left-hand panel. The SFR$_{CO43}$
  assuming both BzK and ULIRG SLEDs is shown, as in
  Fig.\ref{fig:870vsco}. Gray squares show for comparison the SFR as
  determined from L$_{IR,870\mu m}$ with a MS SED (the dashed line at
  the top left shows the effect of using a SB SED). Empty symbols show
  the {\it ALMA}-detected sources A1 and A2 (see text).  Arrows show
  $2\sigma$ upper limits as in Fig.~\ref{fig:870vsco}. The black and
  gray error bars at the top left show the uncertainties due to the
  scatter in the adopted scaling relations for L$_{IR,CO43}$ and
  L$_{IR,870\mu m}$, respectively (see Fig.~\ref{fig:870vsco}). The
  black error bar at the bottom right shows the typical uncertainty on
  stellar masses.  The black thick and thin lines show the MS at the
  cluster redshift \citep[parametrised as in][]{sargent2014} and its
  estimated ($1\sigma$) scatter. The horizontal lines show the
  3$\sigma$ limits on SFRs inferred from the 870$\mu$m and CO(4-3)
  line emission, as indicated (see text). The vertical line shows the
  mass completeness limit of the S16 sample, given the F140W magnitude
  threshold m140=24.7, assuming a solar metallicity SSP with no dust
  attenuation formed at $z\sim7$.
\label{fig:almaRS}}
\end{figure*}

Despite the poor statistics due to the very small field probed and
relatively small number of massive star-forming cluster members, the
{\it ALMA}-detected sample in this region appears unusually skewed
towards very red (F105W-F140W, dust-uncorrected) sources. We show in
Figure~\ref{fig:sfrvscolors} (bottom panel) the color distribution of
sub-samples of the {\it ALMA}-detected sources in ClJ1449 with
different stellar mass and SFR thresholds, as indicated. The adopted
stellar mass thresholds log(M/M$_{\odot}$)=10.1 and 10.5 correspond to
the lowest mass of the {\it ALMA}-detected cluster members, and to a
mass above the mass completeness limit of the S16 sample where our
formal $3\sigma$ limit on CO(4-3)-based SFR probes essentially all the
$1\sigma$ range of the MS (Fig.~\ref{fig:almaRS}). The adopted SFR
thresholds correspond to the nominal $3,5\sigma$ limits of the CO(4-3)
observations in the assumptions discussed in Sec.~\ref{sec:dustySF}.

Figure ~\ref{fig:sfrvscolors} (top panel) shows for comparison the
color distribution of field galaxies at $z_{phot}=2\pm0.3$ with the
same stellar mass and SFR limits, from a control field in GOODS-S. The
control field is the same as used in S16, and the adopted measurements
are described there in mode detail (e.g. their Sec.~2). We briefly
remind here that we used the \citet{guo2013} photometry, and stellar
masses, photometric redshifts and model SEDs from
\citet{schreiber2015,pannella2015}. We note that both stellar mass and
SFR estimates are derived from SED fitting for the field samples,
while we use the CO(4-3)-based SFRs for cluster galaxies. The
selection of the field comparison samples cannot thus be considered as
properly equivalent to the selection of the cluster samples, because
of the initial field sample selection \citep{guo2013} and of the
obvious biases between the different (CO(4-3) vs.\ SED-based) SFR
estimates adopted. In this respect, we further note that for the
purpose of estimating SFRs for the field sample, a constant star
formation history was assumed \citep[model SEDs synthesised
  with][]{bc03} for all sources, allowing for a wide dust attenuation
range ($A_v$=0-6, assuming the \citet{calzetti2000} attenuation law).
SED-based SFRs derived from this modeling have been shown to agree
with L$_{IR}$-based estimates within a $\sim$0.2~dex scatter
\citep[e.g.][for the same SED analysis as used here]{pannella2015}.

The green and blue/orange histograms in Fig.~\ref{fig:sfrvscolors}
(top panel) refer to, respectively, samples including all galaxies or
only galaxies classified as star-forming based on their restframe UVJ
colors \citep{williams2009}. While the blue/orange histograms are thus
the main reference for the expected color distribution, the green
histograms are shown for comparison to account for misclassification
of dusty star-forming galaxies as quiescent sources; as the bulk of
UVJ-quiescent galaxies are expected to be actually quiescent, this is
a conservative comparison sample in this respect, as it maximises the
fraction of red sources.
 
We show Figure~\ref{fig:sfrvscolors} as an indication of our rough
expectations for the colors of a mass and SFR-selected sample of
(field) galaxies at the cluster redshift. Considering 1000
realizations of galaxy samples of the same size as the cluster samples
shown in Fig.~\ref{fig:sfrvscolors}, randomly drawn from the
corresponding field distribution, returns a fraction of
F105-F140$>$1.3 sources larger than what measured in the cluster
samples in $<2\%$ of the realizations at worse (for the
log(M/M$_{\odot}$)$>$10.1, SFR$>$60~M$_{\odot}$yr$^{-1}$, or generally
$<0.5\%$)\footnote{Even considering all the UVJ-quiescent sources as
  misclassified dusty star-forming galaxies, the probability to
  observe red fractions as high as in the cluster samples in the
  corresponding samples drawn from the field distributions remains
  $<2\%$ (or $<7\%$ for the log(M/M$_{\odot}$)$>$10.1,
  SFR$>$60~M$_{\odot}$yr$^{-1}$ sample).}.  In spite of the caveats
outlined above deriving from the non-equivalent selection of the
cluster and field samples due to the different adopted SFR indicators,
and of the very small number statistics of the {\it ALMA}-detected
sample in the cluster, the comparison with the first-order
expectations from the field sample suggests at face value that very
obscured sources are more prevalent than in the field. If confirmed,
this would point towards environmental effects possibly related to
merger-driven star formation episodes (see e.g. C18 and references
therein) and/or differences in star formation histories in the cluster
environment.

\subsection{The {\it ALMA} view of the red cluster galaxy population}
\label{sec:quiescent}

In S16 we investigated the optically red (F105-F140$\gtrsim$1) cluster
galaxy population trying to define its nature, thus probing quenching
of star formation and the early appearence of the red sequence in
dense environments. Note that this red sample is not a {\it red
  sequence} sample (see S16), as it includes galaxies bluer or redder
than the red sequence that are redder than the bulk of the blue
star-forming cluster galaxy population. As discussed in S16, cluster
galaxies with red (dust-uncorrected) optical colors (F105-F140,
$\sim$U-B restframe) can obviously be quiescent sources with evolved
stellar populations, quenching galaxies with still relatively young
stellar populations, as well as massive dusty star-forming
galaxies. Given the characteristics of the available data, in S16 we
adopted a star-forming vs.\ quiescent classification based on two
broad-band colors approximately probing the restframe U-B and B-I
(Fig.\ref{fig:almaRS}--left, see S16).  Although this selection
statistically allows the identification of quiescent vs.\ dusty
star-forming galaxies, it is still likely to be affected by
uncertainties and systematics on an object-by-object basis.

The {\it ALMA} observations presented here offer a fully independent
view of the red population.  Figure~\ref{fig:almaRS} shows the {\it
  ALMA} constraints on SFR vs.\ stellar mass for the S16 red
sample. The stellar masses shown for this sample are generally
SED-based estimates from S16, with the exception of HN7 and HS7 for
which most of the photometry used in S16 is expected to be affected by
contamination given their very small angular separation. We thus use
here stellar mass estimates derived from just the high-resolution
F105W and F140W photometry, by empirical calibration on a sample of
galaxies in a similar magnitude and redshift range from the GOODS-S
control field (as introduced in Sec.~\ref{sec:colordist}). SFR
estimates are derived from the 870$\mu$m continuum and CO(4-3) line
emission as discussed in Sec.~\ref{sec:dustySF}.
Figure~\ref{fig:almaRS} shows for reference (horizontal lines) the
3$\sigma$ limits on such SFRs corresponding to the limits on
L$_{IR,CO43}$ and L$_{IR, 870\mu m}$ discussed in
Section~\ref{sec:dustySF}, Figure~\ref{fig:870vsco}. Also here, these
are shown as an indication, while the measurements, errors and upper
limits for the individual sources account for the position within the
band~4 primary beam, and for the systematics due to the SED or CO SLED
choice (upper limits for SFR$_{CO(4-3)}$ still assume a line width of
400~km/s).

Given the depth of the {\it ALMA} observations, the SFR vs.\ stellar
mass constraints are mostly effective at high stellar
masses\footnote{Less massive sources, e.g. H8, H9, H12, would not be
  detected in our observations even if they were MS star-forming
  galaxies.}, log(M/M$_{\odot}$)$>$10.5. In this mass range, SFR upper
limits suggest a confirmation of the quiescent picture for most
sources color-classified as passive, H2, H4 and H5\footnote{H2 might
  have a tentative, low-significance 870$\mu$m detection with no
  associated CO(4-3) emission (see C18 and Fig.~\ref{fig:almaRS})
  placing it anyway below the MS scatter.}. The only exception is H3,
which is color-classified as quiescent, it is undetected at 870$\mu$m,
but has a 4.8$\sigma$ CO(4-3) detection. Given the estimated
uncertainties and scatter in the adopted scaling relations
(Figs.~\ref{fig:870vsco}, ~\ref{fig:almaRS}) the measurements are
still consistent with a MS SFR for this source. On the other hand,
given the complex morphology and surroundings of this galaxy, and the
limited resolution of our {\it ALMA} observations, it is possible that
the ``quiescent colors'' and the CO(4-3) line belong to different
components.

Concerning instead cluster members in this ``red'' sample
color-classified as star-forming, they are all (but see H7 below)
detected at both 870$\mu$m and CO(4-3). Note that the population of
lower-mass (log(M/M$_{\odot}$)$\lesssim$10.3) blue star-forming
cluster galaxies \citep[][S16]{gobat2013} is not in general expected
to be detected (even when falling within the probed field of view)
given the SFR limits shown in Fig.~\ref{fig:almaRS}. The AGN host
\citep{gobat2013} H13 is consistent with being a MS star-forming
galaxy. H1 and H6 had a particularly uncertain color classification
(see Fig.\ref{fig:almaRS}--left and S16), and indeed H6 was classified
as quiescent based on previous SED \citep{strazzullo2013} and spectral
\citep{gobat2013} modeling.  Although we are investigating other
interpretations for H6 \citep[e.g.][]{gobat2017} involving a quiescent
component, its SFR estimates from the observed 870$\mu$m continuum and
CO(4-3) lime emission are fully consistent with each other, and in
line with MS expectations, suggesting that H6 is likely a MS
galaxy. On the other hand, Fig.~\ref{fig:almaRS} suggests that the SFR
of H1 might be below the MS scatter. Also note that, although we do
our best in dealing with close neighbours by modeling such sources
simultaneously when measuring both 870$\mu$m and CO(4-3) fluxes (see
C18), the continuum and line fluxes of H1 are likely affected (to
different extents) by residual contamination from neighbouring sources
and in particular the very bright A1 (see Fig.~\ref{fig:cutouts}),
which might possibly explain also the relatively large L$_{IR,870\mu
  m}$ vs.\ L$_{IR,CO43}$ offset for this source in
Fig.~\ref{fig:870vsco}.

\begin{figure*}
\begin{center}
\includegraphics[width=0.9\textwidth,bb=0 77 961 714,clip  ]{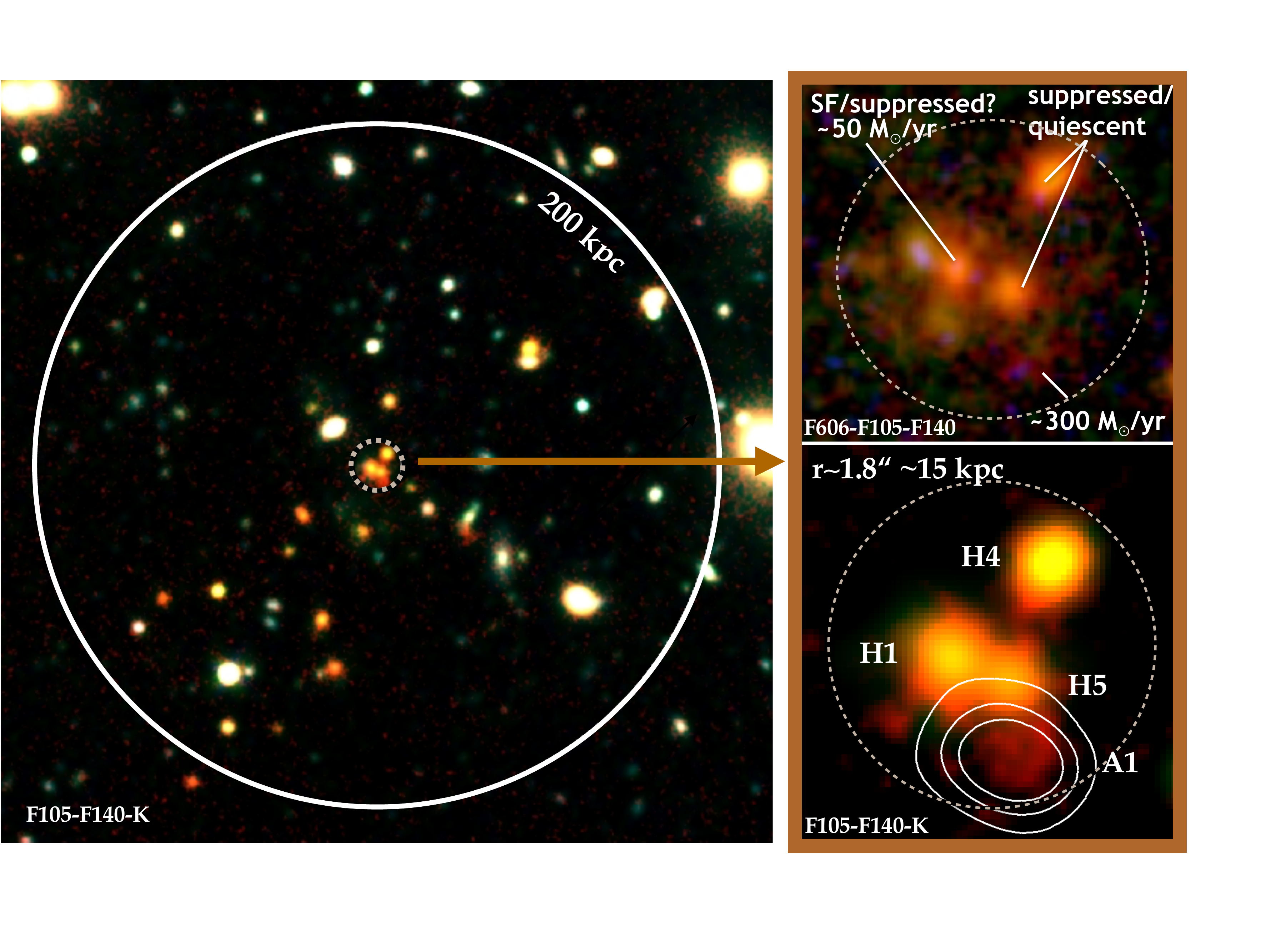}
\end{center}
\caption{The forming BCG at the center of ClJ1449. {\it Left:} A
  F105-F140-K ({\it HST}, Subaru) pseudo-color image of the cluster core
  ($r\lesssim200$~kpc proper). {\it Right:} F606-F105-F140 (top) and
  F105-F140-K (bottom, with overlaid $3,6,9\sigma$ 870$\mu$m contours)
  pseudo-color images of the central $r\lesssim15$~kpc. The dashed
  circle ($r=1.8''\sim15$~kpc) encompasses three F140W-bright
  optically-red sources (H1, H4, H5), the brightest CO(4-3) and
  870$\mu$m cluster member (A1), and several other (unmarked) faint
  diffuse components. The labels assigned to the four highlighted IDs
  in the top-right panel refer to the characteristics discussed in
  Secs.~\ref{sec:dustySF}, \ref{sec:quiescent}, \ref{sec:bcg}.
  \label{fig:bcg}}
\end{figure*}

Figure~\ref{fig:almaRS} (right) also shows the two components HN7 and
HS7 of a very close pair hosting an AGN \citep{gobat2013}. In
Figures~\ref{fig:870vsco} and \ref{fig:almaRS} (left, S16) they are
shown as a single source H7 because the angular resolution attained
does not permit deblending of the two components. Given the shape and
size of the synthesised 870$\mu$m beam, the low S/N of the 870$\mu$m
detection, and the angular distance between the two components, it is
not possible to measure the 870$\mu$m flux of the two components
separately. On the other hand, the position of the higher-S/N CO(4-3)
detection could suggest that HN7 provides the dominant contribution to
the observed CO emission (see HN7-related panels in
Fig.~\ref{fig:cutouts}). More importantly, given the redshift
difference of the two components, the weak tentative detection of
CO(4-3) line emission of HS7 ($\sim2\sigma$, see C18) has a relatively
large separation in frequency ($\sim0.7$~GHz) from the much stronger
detection of HN7, which allows to separate the respective CO(4-3)
contributions (see C18 for full details). Given that the total
(HN7+HS7) 870$\mu$m-derived IR luminosity L$_{IR,870\mu m}$ is fully
consistent with the CO(4-3)-based estimate L$_{IR,CO43}$ measured at
the position and redshift of HN7 (Fig.~\ref{fig:870vsco}), we conclude
that HN7 provides the largest contribution to the measured 870$\mu$m
flux, with the contamination from HS7 being very likely
negligible. The SFR estimated for HN7 is consistent with the expected
MS level (Fig.~\ref{fig:almaRS}). The faintness of HS7 would suggest a
SFR below the MS level (Fig.~\ref{fig:almaRS}).

Given the faint optical counterparts of A1 and A2 (see discussion in
Sec.~\ref{sec:ovd}), these sources were not part of the S16 red
sample. Their colors (Fig.~\ref{fig:cutouts}) and estimated IR
luminosities (Fig.~\ref{fig:870vsco}) clearly point towards these
being dusty star-forming sources, likely interacting/merging with
their optically-bright close neighbours (see also C18).  These
galaxies are shown as empty symbols in Figure~\ref{fig:almaRS}. We
note that given the faintness and surroundings of both sources we
could not derive reliable stellar masses from multi-band SED fitting
as for the other galaxies in our sample. The stellar mass estimates
shown in the figure for A1 and A2 are derived from dynamical masses
inferred from the CO(4-3) resolved emission and measured line widths,
under the assumptions detailed in C18. As discussed in C18, at least
for A1 this estimate is consistent with a stellar mass derived
empirically from the F105W-F140W colors and F140W magnitudes as
described above.

Figure~\ref{fig:almaRS} thus shows our current best picture of the
massive (log(M/M$_{\odot}$)$>$10.5) cluster galaxy population in the
central region of ClJ1449, highlighting once more \citep[and to a more
  significant degree than our previous work,
  e.g.][]{gobat2011,gobat2013,strazzullo2013,strazzullo2016} its
composite nature combining very actively star-forming galaxies and
sources with already suppressed star formation.

\section{A panchromatic snapshot of a forming Brightest Cluster Galaxy at  \lowercase{$z=2$}}
\label{sec:bcg}

Figure~\ref{fig:bcg} highlights the complex of multiple, likely
interacting components located close to the center of the extended
X-ray emission and galaxy overdensity in ClJ1449, identified as the
forming cluster proto-BCG \citep{gobat2011}, and including sources H1,
H4, H5, and A1. We note that although other cluster members (H2, H6)
have stellar masses consistent with the individual masses of H1, H4,
and H5 (see S16 and Fig.~\ref{fig:almaRS}), the configuration of the
H1, H4, H5, A1 complex discussed here below, and its location with
respect to the galaxy overdensity and X-ray emission, is clearly much
more suggestive, as compared to H2 or H6, of the site of main
formation of the future cluster BCG. We stress for clarity that we
identify this galaxy complex as a whole as the {\it forming
  proto-BCG}, and that we do not observe in ClJ1449 any galaxy already
exhibiting the peculiar features of BCGs.

Within a radius of $r=1.5$'' we identify the two massive quiescent
sources H4 and H5, the massive star-forming galaxy H1 with potentially
sub-MS star-formation activity, and the optically faint and very red,
mm-brightest cluster member A1 which is forming stars at a few
hundred solar masses per year (Secs.~\ref{sec:ovd},
\ref{sec:dustySF}, \ref{sec:quiescent}). Indeed, {\it Spitzer/MIPS}
24$\mu$m imaging already suggested this level of star formation
associated with the proto-BCG \citep[with obvious uncertainties
  related to the probed restframe wavelength and poor angular
  resolution,][]{gobat2011}. The combination of {\it HST} color imaging
and {\it ALMA} observations prove that the bulk of the star formation
in the proto-BCG is actually occurring in the optically-faint,
seemingly minor component A1, rather than in the optically-bright
source H1 as originally thought \citep{gobat2013,strazzullo2013}.
Additional faint components and tails, whose cluster membership and
properties cannot be reliably determined with the current
observations, are observed within $r\lesssim1.5$'' from the proto-BCG
(Fig.~\ref{fig:bcg}).

These observations are strongly suggestive of an actively forming BCG
still assembling its stellar mass through star formation {\it and}
merging. This is in line with results from the few first studies on
distant BCGs, suggesting a strong increase in the fraction of highly
star-forming systems \citep{webb2015b,mcdonald2016a, bonaventura2017},
with $L_{IR}>10^{12}$L$_{\odot}$ sources likely approaching
$\gtrsim50\%$ of BCGs towards $z\sim2$. As also discussed in previous
work \citep{webb2015a,webb2015b,kubo2016,bonaventura2017} gas-rich
mergers might play a significant role in this phase of BCG evolution,
as suggested by our observations as well \citep[see also the related
  discussion on merger-driven star formation in ClJ1449 at large
  in][]{coogan2018}.  Besides the on-going star formation activity, we
stress though that a significant fraction of the stellar mass in the
proto-BCG complex has already evolved to a seemingly quiescent
phase. The depletion timescale estimated in C18 for the most actively
star-forming component A1 is of order 100~Myr. The overall picture is
qualitatively reminiscent of hierarchical model renditions
\citep[e.g.][]{deluciablaizot2007,tonini2012}, though the observations
presented here alone are obviously not sufficient to discuss the
details of such modeling.  Given the estimated (baryonic) masses of A1,
H1, H4, H5 and their relative projected distance, an approximate
estimate of the merger timescale would be of order a few hundred
million years \citep[e.g.][the orbital timescale giving a lower limit
  of $\sim$100~Myr]{lotz2011}, with an estimated total stellar mass of
the resulting BCG $\sim3\times10^{11}$M$_{\odot}$.

\section{Summary}
\label{sec:conclusions}

{\it ALMA} observations of the 870$\mu$m continuum and CO(4-3) line
emission in the central region of ClJ1449 have significantly improved
our understanding of galaxy populations in this cluster
core. Crucially, CO(4-3) follow-up secured spectroscopic confirmation
of optically faint, mm-bright cluster members, while questioning the
membership of the two brightest 870$\mu$m sources in the field.

 The 870$\mu$m continuum and CO(4-3) line emission yield a total
estimated SFR within the probed $\sim0.08$~Mpc$^2$ region of
$\sim700\pm100$~M$_{\odot}$/yr, resulting in a projected SFR density
of $\sim0.9\pm0.1\times10^{4}$~M$_{\odot}$~yr$^{-1}$Mpc$^{-2}$, and a
SFR volume density five orders of magnitude larger than in the field
at the same redshift. The inferred lower limit (that is, not
correcting for the missing SFR from the portion of the virial volume
not probed by the {\it ALMA} observations) on the SFR density per halo
mass is SFR/M$_{halo}\gtrsim1300\pm$400~M$_{\odot}$~yr$^{-1}$ /
10$^{14}$M$_{\odot}$, which at face value is consistent with
extrapolations from lower redshift observations predicting high SFR
densities in massive (yet sub-$10^{14}$M$_{\odot}$) haloes at this
redshift (see discussion in Sec.~\ref{sec:dustySF}). In spite of its
relatively significant SFR density, the core of ClJ1449 seems
nonetheless far from the $>3000$~M$_{\odot}$~yr$^{-1}$ observed within a
similar (or smaller) clustercentric distance in the similarly massive
ClJ1001+0220 at $z=2.5$ \citep{wang2016}. Three of the five most
massive galaxies (H2, H4, H5, log(M/M$_{\odot}$)$\sim11\pm0.1$) in the core of
ClJ1449 are seemingly quiescent sources remaining undetected at
nominally at least 2$\sigma$ below the Main Sequence (the other two
being the MS star-forming galaxy H6 and the possibly suppressed H1,
see Sec.~\ref{sec:quiescent}, Fig.~\ref{fig:almaRS}), compared to a
star-forming galaxy fraction of 80\% observed at
$M\geq10^{11}$M$_{\odot}$ in the core of ClJ1001+0220
\citep{wang2016}.

The combination with previously available {\it HST} imaging (and grism
spectroscopy) critically enhances the interpretation of these
observations. Although there is generally a close correspondence
between the mm- and optical/NIR-inferred pictures of most cluster
galaxies discussed here, there are particular sources that can only be
really understood by comparing the two. The bright {\it ALMA} sources
A1 and A2 are associated with optical counterparts otherwise deemed
comparatively minor components in the {\it HST} NIR
imaging. Figure~\ref{fig:cutouts} underlines that, if deep
high-resolution color imaging were not available, these {\it ALMA}
detections would appear as dust and gas significantly offset from
stellar emission in the nearest optical/NIR counterpart. The {\it HST}
imaging reveals instead the faint, extremely red components perfectly
matching the sub-mm emission, and probably related to very recent or
ongoing merging events (see further discussion in C18).

The {\it HST} and {\it ALMA} synergy also provides in this cluster
another striking snapshot of the early evolution of forming BCGs
\citep[e.g.][see Sec.~\ref{sec:bcg}]{webb2015a,kubo2016}, with a
seemingly multiple-merger system of quiescent and highly star-forming
components likely assembling the future BCG.

The results based on  {\it ALMA} observations presented here extend the
reach of our previous studies in ClJ1449, drawing quantitative
details in a picture combining, right into the cluster core, a star
formation activity approaching a thousand solar masses per year, the
first massive quiescent cluster-core galaxies, and the ongoing
formation of the BCG through merging of already quiescent and still
vigorously star-forming components.

\acknowledgements Largely based on observations from {\it ALMA}
programs 2012.1.00885.S and 2015.1.01355.S and {\it {\it HST}}
programs GO-11648 and GO-12991. We thank the anonymous referee for a
constructive report improving the presentation of this work. 
RTC acknowledges support from STFC G1687 grant ST/N504452/1.
FV
acknowledges the Villum Fonden research grant 13160 'Gas to stars,
stars to dust: tracing star formation across cosmic time'.
A.R. acknowledges support by an INAF grant PRIN-2017.  N.A. is
supported by the Brain Pool Program, which is funded by the Ministry
of Science and ICT through the National Research Foundation of Korea
(2018H1D3A2000902). AC acknowledges PRIN MIUR 2015 "Cosmology and
Fundamental Physics: illuminating the Dark Universe with
Euclid''. H.D. acknowledges financial support from the Spanish
Ministry of Economy and Competitiveness (MINECO) under the 2014
Ram{\'o}n y Cajal program MINECO RYC-2014-15686.

\bibliographystyle{aa}

\end{document}